# A Detailed Assessment of Numerical Flow Analysis (NFA) to Predict the Hydrodynamics of a Deep-V Planing Hull

T. C. Fu[1], T. T. O'Shea[2], C. Q. Judge[3], D. Dommermuth[2],
K. Brucker[2], and D. C. Wyatt[2]
([1]Naval Surface Warfare Center, Carderock Division, [2]Science Applications
International Corporation, [3]United States Naval Academy, USA)

**ABSTRACT**

Over the past few years much progress has been made in Computational Fluid Dynamics (CFD) in its ability to accurately simulate the hydrodynamics associated with a deep-V monohull planing craft. This work has focused on not only predicting the hydrodynamic forces and moments, but also the complex multiphase free-surface flow field generated by a deep-V monohull planing boat at high Froude numbers. One of these state of the art CFD codes is Numerical Flow Analysis (NFA). NFA provides turnkey capabilities to model breaking waves around a ship, including both plunging and spilling breaking waves, the formation of spray, and the entrainment of air. NFA uses a Cartesian-grid formulation with immersed body and volume-of-fluid methods.

The focus of this paper is to describe and document a recent effort to assess NFA for the prediction of deep-V planing craft hydrodynamic forces and moments and evaluate how well it models the complex multiphase flows associated with high Froude number flows, specifically the formation of the spray sheet. This detailed validation effort was composed of three parts. The first part focused on assessing NFA's ability to predict pressures on the surface of a 10 degree deadrise wedge during impact with an undisturbed free surface. Detailed comparisons to pressure gauges are presented here for two different drop heights, 15.24 cm (6 in) and 25.4 cm (10 in). Results show NFA accurately predicted pressures during the slamming event. The second part examines NFA's ability to match sinkage, trim and resistance from Fridsma's experiments performed on constant deadrise planing hulls. Simulations were performed on two 20 degree deadrise hullforms of varying length to beam ratios (4 and 5) over a range of speed-length ratios (2, 3, 4, 5, and 6). Results show good agreement with experimentally measured values, as well as values calculated using Savitsky's parametric equations. The final part of the validation study focused on assessing how well NFA was able to accurately model the complex multiphase flow associated with high Froude number flows, specifically the formation of the spray sheet. NFA simulations of a planing hull fixed at various angles of roll (0, 10, 20, and 30 degrees) were compared to experiments. Comparisons to underwater photographs illustrate NFA's ability to model the formation of the spray sheet and the free surface turbulence associated with planing boat hydrodynamics. Overall these three validation studies provide a detailed assessment on the current capabilities of NFA to predict the hydrodynamics of a deep-V planing hull.

**INTRODUCTION**

As interest in high-seed craft has grown in recent years the number of efforts to experimentally characterize and numerically simulate their behavior has also increased. Due to the reduced submergence of high-speed planing craft, small motions can result in large changes in wetted surface area, and linearity assumptions are less valid since hydrodynamic forces dominate in the planing regime making these problems especially challenging. Building upon the classic prismatic planing hull experimental efforts of Fridsma (1969) and Savistsky (1964), a number of researchers have focused on the problem of dynamic stability of high speed planing craft. Codega and Lewis (1987) described a class of high-speed planing boats that exhibited dynamic instabilities such as the craft trimming by the bow, rolling to a large angle of heel to port, and broaching violently to starboard. Blount and Codega (1992) presented data on boats that exhibited non-oscillatory dynamic instabilities and suggested quantitative criteria for development of dynamically stable planing boats. Katayama et al. (2007) found that instability is strongly influenced by the running attitude of the hull at high speeds. Troesch 1992, Troesch & Falzarano 1993 have investigated the dynamic behavior of planing, and porpoising behavior. Research done at the Stevens Institute of Technology has investigated the transverse stability of planing craft using coupled linear sway/roll/yaw equations of motion (Lewandowski 1997).

Significant progress has been made in Computational Fluid Dynamics (CFD) and its ability to accurately simulate the hydrodynamics associated with a deep-V monohull planing craft, see (Fu et al, 2010; Broglio & Iafarati, 2010; Fu et al 2011, Jiang 2012). This current validation effort is focused on predicting the hydrodynamic forces and moments, but also the complex multiphase free-surface flow field generated by a deep-V monohull planing boat at high Froude numbers. One of these state of the art CFD codes is Numerical Flow Analysis (NFA). NFA provides turnkey capabilities to model breaking waves around a ship, including both plunging and spilling breaking waves, the formation of spray, and the entrainment of air. NFA uses a Cartesian-grid formulation with immersed body and volume-of-fluid (VOF) methods. O'Shea et al (2008) describes the code and recent applications to naval problems.

This paper describes a recent effort to assess NFA for the prediction of deepv-V planing craft hydrodynamic forces and moments and evaluate how well it models the complex multiphase flows associated with high Froude number flows, specifically the formation of the spray sheet. This detailed validation effort was composed of three parts. The first part focused on assessing NFA's ability to predict pressures on the surface of a 10 degree deadrise wedge during impact with an undisturbed free surface. The second part examines NFA's ability to match sinkage, trim and resistance from experiments performed on constant deadrise planing hulls, detailed in Fridsma (1969), and the final part of the validation study is focused on assessing how well NFA was able to accurately model the complex multiphase flow associated with high Froude number flows, specifically the formation of the spray sheet.

**NUMERICAL APPROACH**

The Numerical Flow Analysis (NFA) computer code is employed in this study. NFA solves the Navier-Stokes equations utilizing a cut-cell, Cartesian-grid formulation with interface-capturing to model the unsteady flow of air and water around moving bodies. The interface-capturing of the free surface uses a second-order accurate, volume-of-fluid technique. NFA uses an implicit subgrid-scale model that is built into the treatment of the convective terms in the momentum equations (Rottman, et al., 2010). A panellized surface representation of the ship hull (body) is all that is required as input in terms of body geometry. Domain decomposition is used to distribute portions of the grid over a large number of processors. The algorithm is implemented on parallel computers using FORTRAN 2003 and the Message Passing Interface (MPI). The interested reader is referred to Dommermuth, et al., (2007), O'Shea, et al., (2008), and Brucker, et al., (2010) for a detailed description of the numerical algorithm and of its implementation on distributed memory high performance computing (HPC) platforms.

**VALIDATION PART 1: WEDGE DROP**

The structural response of a planing boat during a slamming event is of interest to designers. Specifically the fatigue life of the structural members of existing planing boat designs are observed to be shorter than would be expected given current design tools. During a slamming event the bottom of a planing hull experiences a severe pressure pulse as the spray root moves outboard from the keel. To quantify this high speed high pressure pulse a wedge drop experiment was performed at the Naval Surface Warfare Center, Carderock Division (NSWCCD), the details of which can be found in Lesar et al (2012) and Jiang et al (2012). The wedge was instrumented with, among other things, an array of high frequency pressure gauges. During the impact these gauges captured the pressure over time at various distances from the keel.

To assess NFA's ability to accurately predict these pressures, simulations of the experiments were preformed. Care was taken to insure the initial conditions and boundary conditions were matched to the experimental setup. The wedge was dropped from rest at 15.24 cm (6 in) and 25.4 cm (10 in) above the undisturbed free surface. The model tank and the NFA domain were 4.6 m (15 ft) wide by 7.3 m (24 ft) long by 4.4 m (14 ft 6 in) deep. The wedge weighed 75 kg (165.2 lbs) and was 91.4 cm (36 in) by 0.60 cm (23.75 in) with a deadrise angle of 10.23 degrees. Figure 1 shows the bottom of the instrumented wedge alongside the NFA simulation's geometry and domain extents. Two symmetric pressure probes were located at 3.8 cm (1.5 in) from the keel. A four by four array of pressure probes had lines of 4 probes at 13.68, 17.26, 20.84, and 24.42 cm (5.385, 6.795, 8.205, and 9.615 in) from the keel. Each probe in the array was 3.58 cm (1.41 in) from its neighbor longitudinally and transversely. For both drop heights the NFA simulations used wall boundary conditions in each Cartesian direction. The number of grid points in the x, y and z directions was 1536, 1024, and 512, resulting in 805 million cells. Grid spacing near the wedge was 0.001L or 0.09 cm (0.035 in), where L is the wedge length. The nondimensional time step taken was 0.00005, resulting in an output frequency of 37.8 KHz. The simulations were run on Jade, a Cray XT5, on 768 processors and took 12 hours for the 15.24 cm (6 in) drop height, running for 2000 time steps, and 32 hours for the 25.4 cm (10 in) case, running 6000 time steps.

The output from the pressure probes is plotted against the pressure sampled from NFA in Figures 2 and 5 for the 6 15.24 cm (6 in) and 25.4 cm (10 in) drop heights respectively. The x axis is time in seconds and the y axis is gauge pressure in psi. Pressure-time histories from three completely separate drops are plotted together, with the peak from the first row of gauges aligned. The dashed black lines show 5 peaks in time, the first from the symmetric keel gauges and the remaining four from the four rows of pressure gauges in the array. Since the pressure pulse is quasi-2D each row of pressure gauges in the keel-wise direction shows approximately the same time history. The time histories exhibit variability associated with minor experimental irregularities. The blue error bars represent the min and max of the experimental peaks for the 3 drops. The sampling frequency of the pressure probes was 19.2 KHz, while NFA pressures were extracted at a rate of 37.8 KHz.

The peak from NFA's first row of gauges was aligned to the peak from the experiments with and plotted in red. NFA shows excellent agreement for peak pressure magnitudes and pulse duration. The spray root takes approximately ten milliseconds to travel across the array of gauges and each peak pressure is maintained for less than 1ms. The propagation in time of the pressure pulse from the spray root is slowed by the deceleration of the wedge due to the impact with the free surface. The length of time between the final two peaks from is therefore greater than that of the preceding peaks. NFA correctly predicts this momentum transfer and thus the position in time of all four pressure peaks. NFA is within experimental variability for peak pressure and timing. It is notable that NFA also predicts the brief negative pressures seen in the experiments before the peak pressure due to the turbulent breakup of the expanding spray root.

Figures 3 and 4 show image sequences from the 15.24 cm (6 in) and 25.4 cm (10 in) drop height simulations respectively. The camera is looking up at the wedge from underneath the water surface. The wedge is painted with contours representing the pressure that NFA has calculated. Blue represents zero pressure while red indicates 103.4 kilopascals (15 psi). The pressure pulse can be seen moving outboard from the keel over the pressure probes indicated by white dots on the bottom of the wedge on the right side. This pulse is concentrated at the spray root which can be seen as the translucent surface that begins to break up later in the simulation.

.

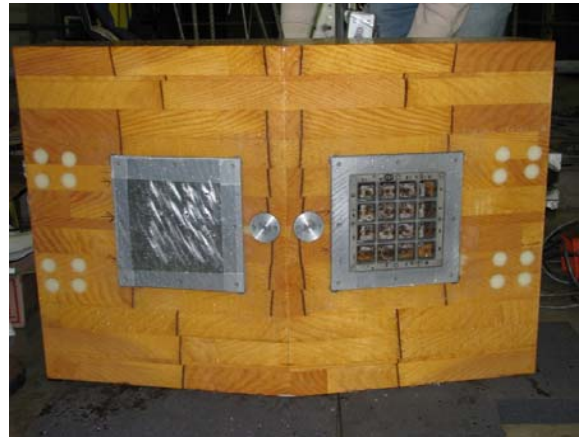
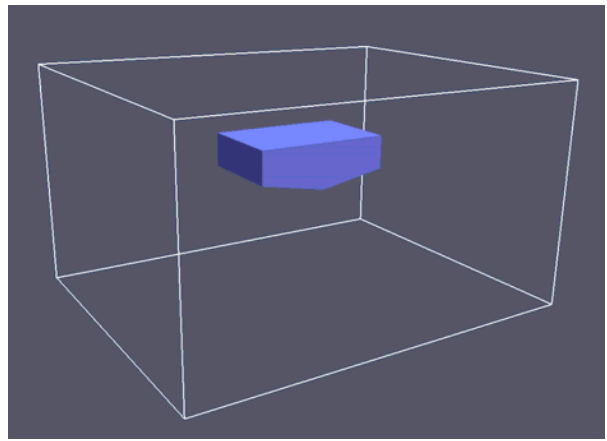

**Figure 1:** Photograph of instrumented wedge used in experiment on the left and NFA wedge geometry and domain size on the right

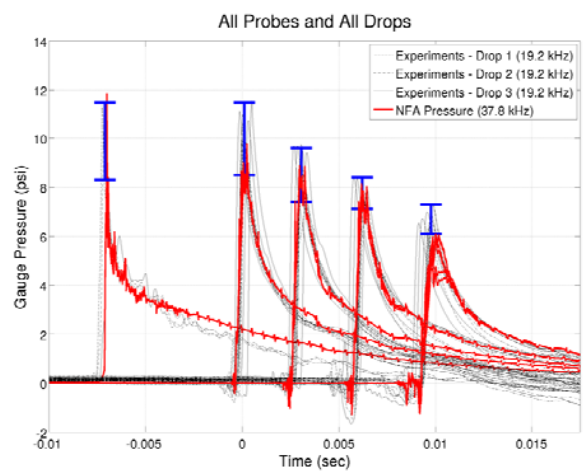

**Figure 2:** 15.24 cm (6 in) drop height case - NFA pressure predictions plotted on top of pressure gauge output from 3 separate drops.

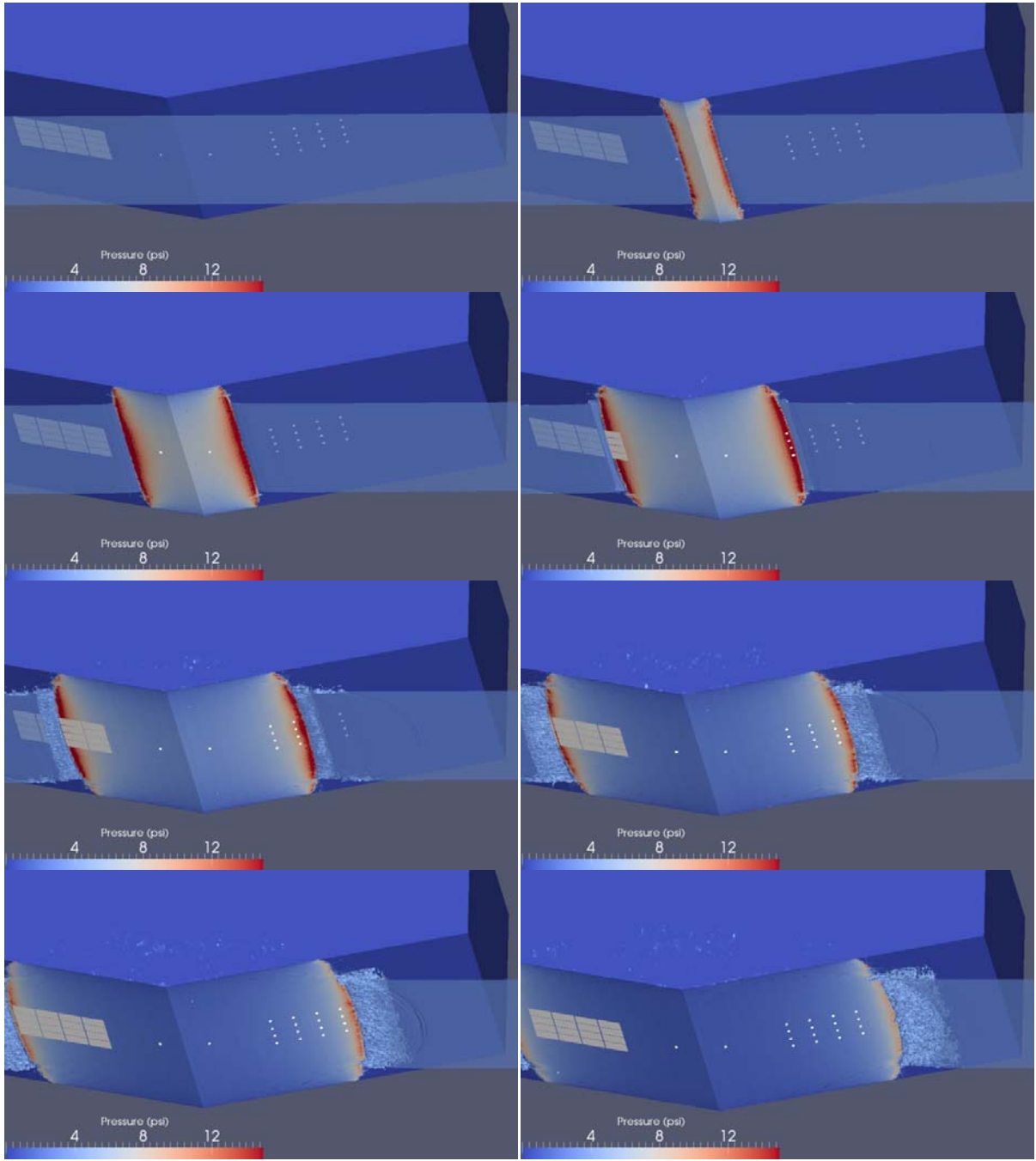

**Figure 3:** Sequences of frames from an animation of the 15.24 cm (6 in) drop height wedge impacting the free surface. Pressure is interpolated onto the wedge and displayed, blue indicating 0 pressure, and red indicating 103.4 kilopascals (15 psi). Note the pressure wave from the spray root expanding transversely over the pressure gauges.

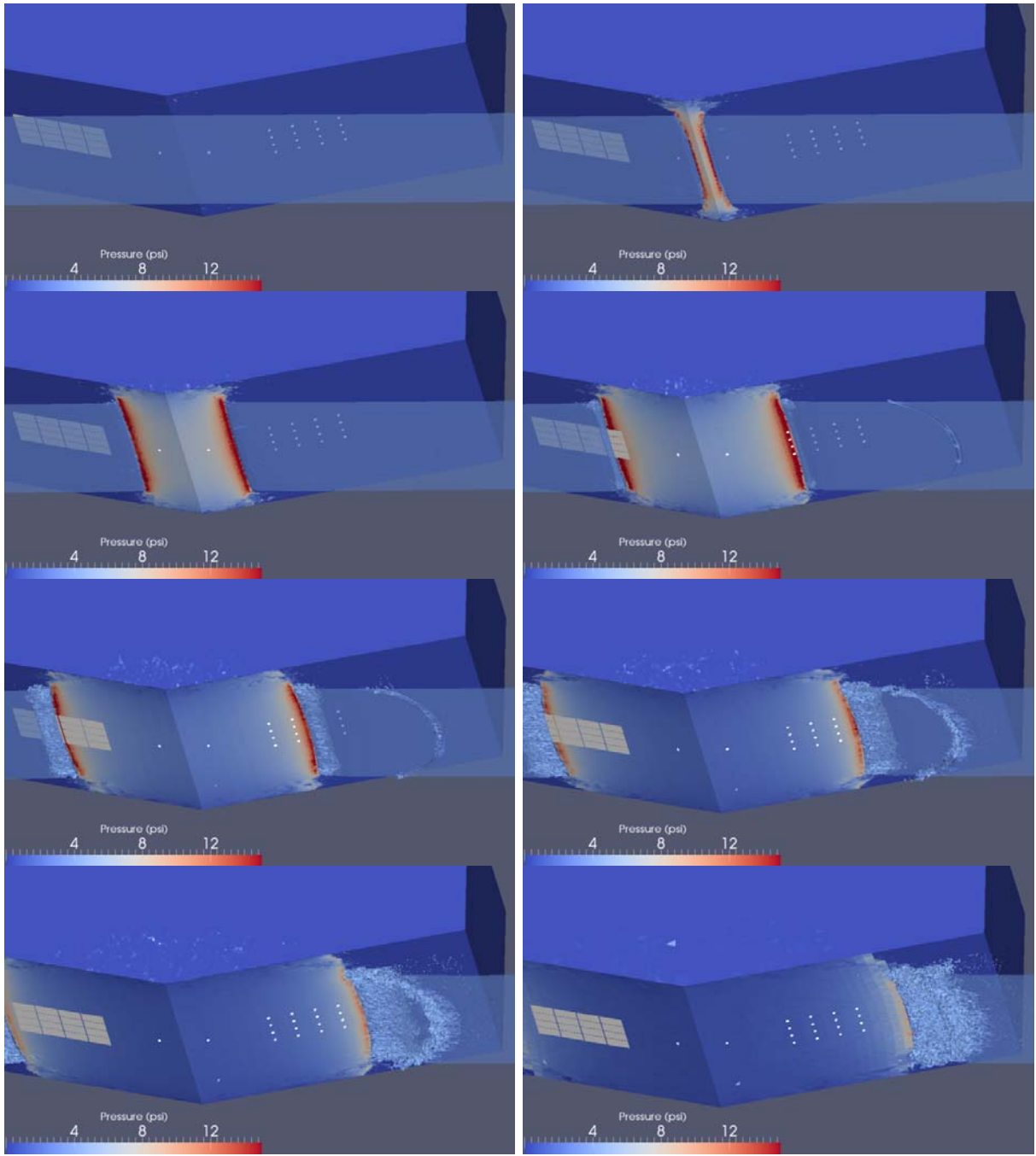

**Figure 4:** Sequences of frames from an animation of the 25.4 cm (10 in) drop height wedge impacting the free surface. Pressure is interpolated onto the wedge and displayed, blue indicating 0 pressure, and red indicating 103.4 kilopascals (15 psi). Note the pressure wave from the spray root expanding transversely over the pressure gauges.

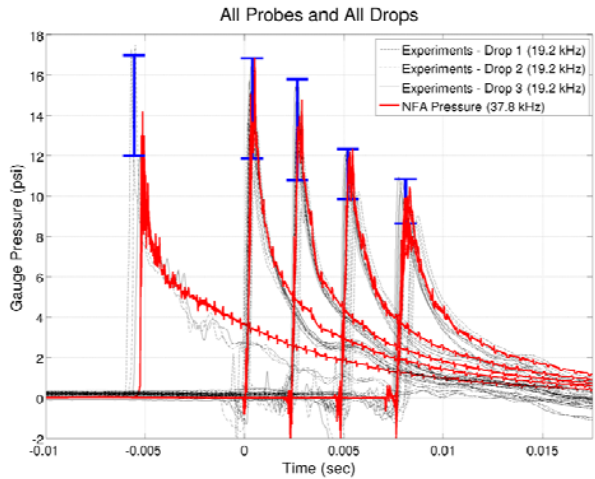

**Figure 5:** 25.4 cm (10 inch) drop height case - NFA pressure predictions plotted on top of pressure gauge output from 3 separate drops.

## VALIDATION PART 2: CONSTAND DEADRISE, STEADY FORWARD SPEED

The planing hull model used in the experiments of Fridsma (1969) is employed for this study. Three configurations with different length (L) to beam (b) ratios, L/b=4 and L/b=5 (shown in Figure 6 (top)), and light/heavy displacements are investigated. In all the simulations the model is of unit length with the bow of the model is fixed at x=0, and the flow is from positive x. The model was free to sink and trim about the center of gravity (CG). The vertical center of gravity (VCG) was fixed at 0.294b above the keel, while the lateral center of gravity (LCG) was fixed at 0.6, 0.655, or 0.7 ship lengths from the bow. Inflow and outflow boundary conditions are used in the stream-wise (x) direction, and free-slip conditions are used in the span-wise (y) and cross-stream (z) directions. The inflow condition is a free-stream current, and the outflow condition is one-dimensional, non-reflective, Orlanski-type boundary condition. The Froude number is Fr=U(gL)-1/2, where U and L are respectively the speed and length of the planing hull model, and g is the acceration of gravity, so Fr varies between 0.6 and 1.8 . The extents of the computational domain, shown in Figure 6 (bottom), in the stream-wise (x), span-wise (y), and cross-stream (z) directions are respectively [-3.0L, 1.0L], [-1.6L, 1.6L], and [-1.25L, 1.0L]. These dimensions were chosen to match the Davidson Laboratory's Tank 3 dimensions (Bruno 1993). The number of grid points [nx, ny, nz] are [768, 512, 384]. The grid is stretched with nearly uniform spacing around the hull where the grid spacing is [0.00304L, 0.00306L, 0.00305L]. The maximum grid spacing far away from the ship along the Cartesian axes is [0.016L, 0.0204L, 0.0152L]. The grid points are distributed in 64x64x64 blocks over 576 cores, and the time step is Δt=0.0005. All simulations have been run on the Cray® (Cray Inc.) XE6, Raptor, platform located at the U.S. Army Engineering Research and Development Center (ERDC) and run by the Air Force Research Laboratory (AFRL).

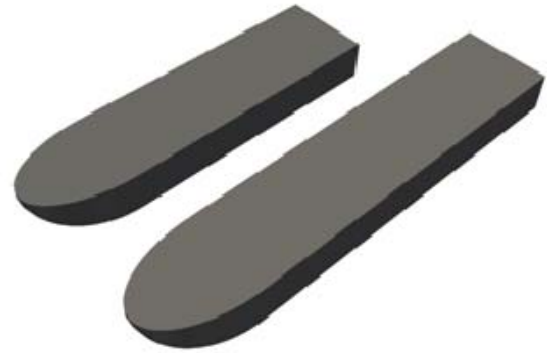

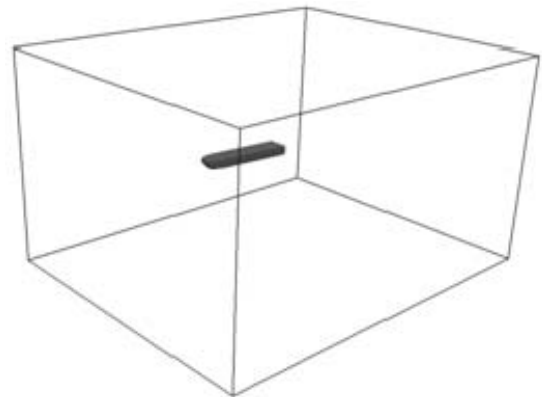

**Figure 6:** Setup L/b=4 and L/b=5 hull geometries (left) and L/b=5 model in computational domain (right).

Figures 7-9 are comparisons of the rise at the CG, or sinkage, on the left, the trim in the middle, and the ratio of the resistance to the displacement on the right. A ▲ denotes the experimental measurements made by Fridsma (1969); the ● denotes the predicted value according to Savitsky (1964); and the ■ denotes the value predicted by the NFA Simulations. The NFA simulations were averaged from T=4-8 for Figures 7 and 8 and over the last two boat lengths for Figure 9. The error bars show the minimum and maximum values over the averaging period for the rise at the CG and trim, and ± the r.m.s. for the resistance. The comparison of interest here is between the experiments and the NFA simulations, the Savitsky predictions are simply provided for completeness.

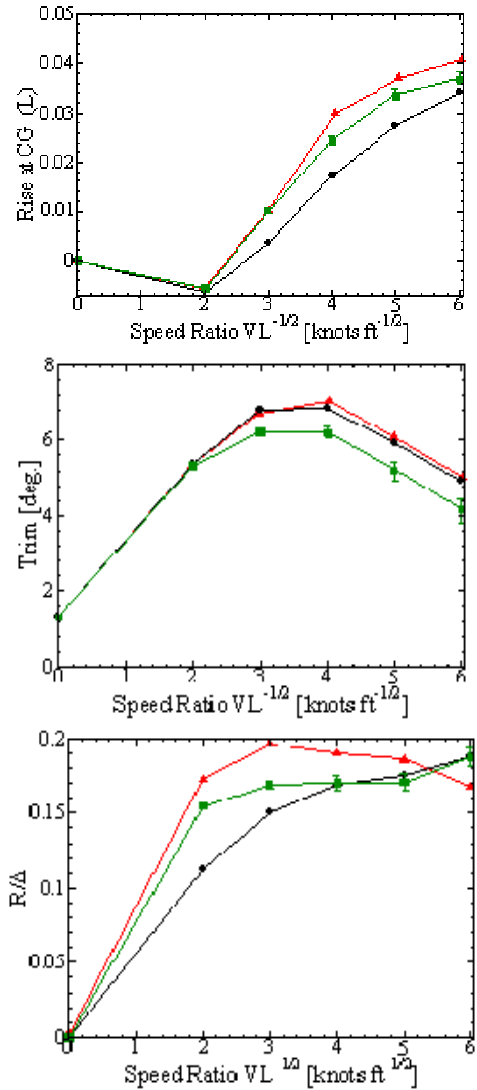

**Figure 7:** Fridsma comparison for $C_\Delta$=0.608, L/b=4 and β=20°. ▲, Fridsma Experiments (Fridsma1969); ●, Savitsky Predictions (Savitsky 1964); ■, NFA Simulations. Rise at the CG (top), trim (middle), and resistance (bottom).

Figure 7 shows the results of the comparison for $C_\Delta$=0.608, L/b=4 and β=20°, with the LCG at 0.6L from the bow. Where $C_\Delta$ is the load coefficient defined as $C_\Delta = \Delta/(wb3)$, where Δ is the hull displacement, w the specific weight of water, b is the beam of the hull, L is the length of the hull, and β is the deadrise angle. The rise at the CG, shown in hull lengths, is shown (top) at the slower speed ratios, $VL^{-1/2}$ [knots-ft$^{-1/2}$], namely two and three, the results of the simulations are identical to those recorded during the experiments. At higher speed ratios, namely four, five and six, the rise at the CG is under predicted by the simulations. The trim, shown in the middle frame, also shows good agreement with the experiments, although again at the higher speeds, the simulations under predict the final trim angle by around ten percent. It is noted that the general shape of the curve matches the shape of the curve recorded in the experiments and also the predictions from Savitsky. The resistance (bottom frame) is under predicted for all but the highest speed case, where it is over predicted. In all three frames of Figure 8 the error bars are small, meaning that the simulations predicted a nearly constant value for the quantities of interest. No porpoising was observed.

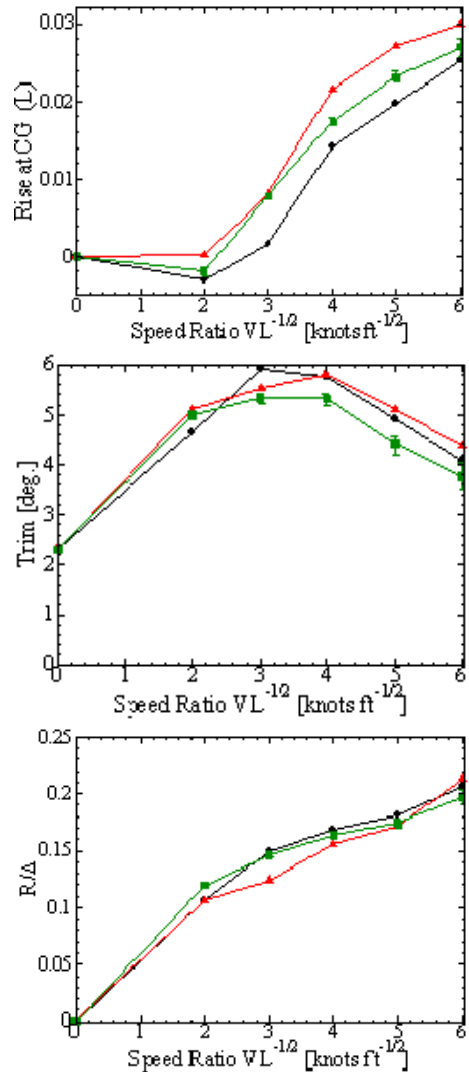

**Figure 8:** Fridsma comparison for $C_\Delta$=0.608, L/b=5 and β=20°. ▲, Fridsma Experiments (Fridsma1969); ●, Savitsky Predictions (Savitsky 1964); ■, NFA Simulations. Rise at the CG (top), trim (middle), and resistance (bottom).

Figure 8 shows the results of the comparison for $C_\Delta$=0.608, L/b=5 and β=20°, with the LCG at 0.655L from the bow. The rise at the CG, shown on the top, is well predicted by the numerical simulations with the agreement being nearly identical at the lower speeds. There is good qualitative agreement at the higher speeds, but again the numerical simulations under predict the rise at the CG at the higher speeds. The trim, shown in the middle, is well predicted by the numerical simulations, being nearly identical to the experiments for speed ratios of two and three. For the higher speed ratios, those above three, the trim is under predicted but the shape of the curve is similar to the one recorded in the experiments and the one predicted by Savitsky. The resistance, shown on the bottom, displays the opposite trend to the L/b=4 case. That is, it is over predicted by the numerical simulations for speed ratios between two and five, and then under predicted for a speed ratio of six. No porpoising was observed.

Figure 9 shows the lighter displacement case with for $C_\Delta$=0.304, L/b=4 and $\beta$=20°, with the LCG at 0.7L from the bow. The rise at the CG, on the left, is under predicted over the entire speed range. Note the large error bars at the speed ratio of five. This case is the speed at which the model displayed evidence of porpoising in the experiments, and no data was recorded. The trim, shown in the middle, again shows excellent agreement with the value recorded in the experiments for the lowest speed ratio, two, but is under predicted for speed ratios of three and four. The resistance, shown on the right, is nearly identical to the values recorded in the experiments. Porpoising was observed at a speed ratio of five, which was consistent with the experiments.

Figure 10 shows the trim as a function of time. For speed ratios of two and three a steady trim is achieved after only two boat lengths. The speed ratio of four achieves a quasi-steady trim after about five boat lengths. However, the speed ratio of five does not achieve a steady trim and furthermore the amplitude of the oscillations does not decay with time.

Figure 11 shows instantaneous pressures on the bottom of the hull for the $C_\Delta$=0.608, L/b=4 and $\beta$=20°, with the LCG at 0.6L from the bow for speed ratios of two through six moving from left to right. The transition from displacement to planing is evidenced by the sharpening of the spray root and lack of hydrostatic pressure aft of the spray root, which can be observed by comparing the leftmost and rightmost frames. The hot spots are transitory in nature and would not be observed if a time averaged pressures were shown. Given their transitory nature they do not significantly contribute to the force balance which determines the position of the hull.

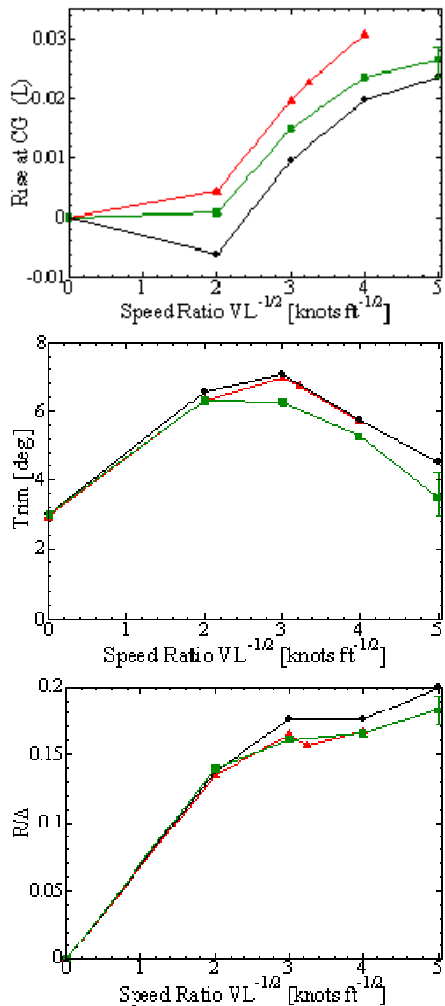

**Figure 9:** Fridsma comparison for $C_\Delta$=0.304, L/b=4 and $\beta$=20°. ▲, Fridsma Experiments (Fridsma1969); ●, Savitsky Predictions (Savitsky 1964); ■, NFA Simulations. Rise at the CG (top), trim (middle), resistance (bottom).

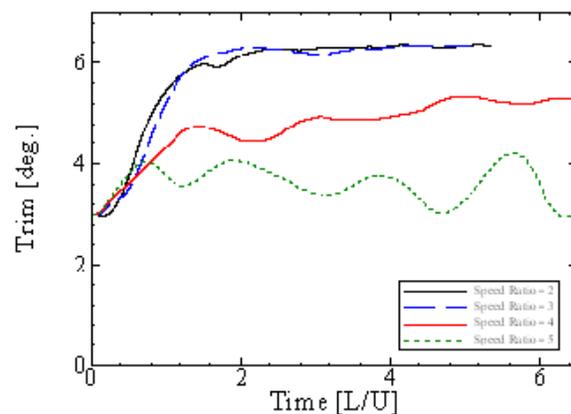

**Figure 10:** Trim versus time for $C_\Delta$=0.304, L/b=4 and $\beta$=20°. Note the oscillatory behavior at a Speed Ratio of 5 (green line), indicative of porpoising. [Speed Ratio is in knots-ft-1/2].

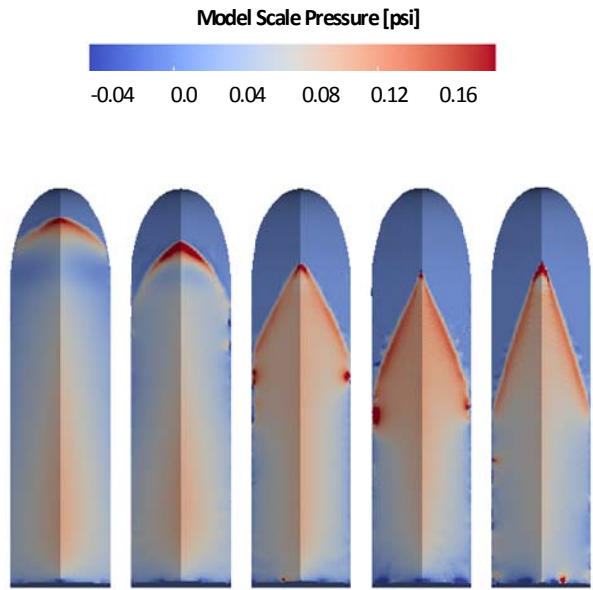

**Figure 11:** Pressures on hull for $C_\Delta$=0.608, L/b=4 and β=20° Speed Ratio = 2, 3, 4, 5, 6 (from left to right).

## VALIDATION PART 3: FIXED ROLL COMPARISONS

In order to investigate the transverse stability and to support the development of CFD codes for high speed, small-craft applications the United States Naval Academy (USNA) performed a series of forced roll experiments on a prismatic planing hull. An experimental mechanism to force a planing hull model in roll motion was designed and built. The Forced Roll Mechanism is intended to act as a dynamometer that forces a planing hull in roll and measures the resulting roll moment as well as the heave and sway forces. The trim and rise of the model can be adjusted between test runs. The dynamometer can be configured to either oscillate the model or hold the model at a fixed roll angle, and measure the forces whether dynamic or static in nature, while allowing different testing conditions in terms of speed, rise and trim. A photograph of the system attached to the wooden 20 degree deadrise model as well as an expanded detail of the forced roll mechanism is shown in Figure 12. The characteristics of the planing hull model are given in Table 1.

With the present motors and instrumentation, the Forced Roll Mechanism (FRM) system is capable of a maximum RPM of 150 (2.5 Hz) and a maximum roll amplitude of 30 degrees for the 20° deadrise wooden planing hull model. The roll angle amplitude is limited by the model clearance with the FRM support structure. Judge (2010) showed that the roll added inertia calculation is highly sensitive to the error in the measured forcing moment amplitude at low frequency. Therefore, to accurately determine the coefficients in roll when oscillating at low frequencies, there must be tighter error control of the forcing moment measurements or an increased number of test runs.

By combining information from the various load cells in the FRM, the total heave and lift forces and roll moment can be determined The block force gauges attached to the supporting heave post assembly measure purely vertical and horizontal (sway) forces in the tow tank reference frame. However, the bi-axial load cell is attached to the model and, therefore, reads vertical and horizontal (side) forces relative to the model. Since the model is trimmed and rotating in roll, these measured forces need to be resolved to find the vertical and horizontal forces relative to the heave post assembly.

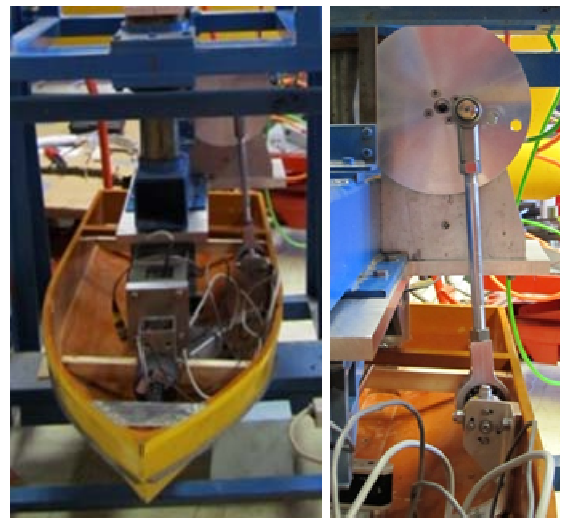

**Figure 12:** Images showing an overview of Forced Roll Mechanism and detail of the roll forcing design.

Testing was performed in the United States Naval Academy Hydromechanics Laboratory's 380 foot tow tank. The USNA tow tank is 116 m (380 ft) long with a width of 7.9 m (26 ft) and a depth of 4.9 m (16 ft). The model was tested at two displacements 14 and 25 kg (30.75 and 55 lbs), three velocities 6.1, 7.6, and 9.1 m/s (20, 25 and 30 ft/s), five fixed roll angles (0, 5, 10, 15, and 20 degrees), three forced roll oscillation amplitudes (10, 15 and 20 degrees), and three forced roll oscillation frequencies (up to 3 Hz). In addition to the roll position of the model, the heave and sway forces and the roll, yaw, and pitch moments during each test were measured (Judge, 2012). The NFA validation effort focused on the first portion of this test, that is, the results for steady forward speed, at fixed sinkage and trim, at a series of different roll angles.

**Table 1:** Characteristics of USNA Planing Hull Model

| | | |
|---|---|---|
| Length on the waterline | 1.524 m (5 ft) | |
| Chine beam | 0.451 m (1.48 ft) | |
| Deadrise | 20 degrees | |
| Displacement | 13.5 kg (29.8 lb) | 26.5 kg (58.4 lb) |
| KG | 13.44 cm (5.29 in) | |
| LCG (fwd transom) | 30.5 cm (12.0 in) | 59.4 cm (23.4 in) |

**Data Analysis**

The lift force, sway force, and roll moment were determined as a function of time for all the steady and dynamic roll tests. The nominal weight of the planing hull was taken to be the lift force from the steady zero-roll test. The fixed trim and heave for the model was determined from the zero-roll equilibrium trim and heave measured during a free to trim and heave run. This weight was used to non-dimensionalize the forces measured. The roll moment was non-dimensionalized by this weight times the beam of the model. The model velocity was non-dimensionalized as the beam Froude number

$$Fr_B = \frac{V}{\sqrt{gB}}.$$

Lift Force

The lift force was calculated from the total forces measured on the model in the direction perpendicular to the water surface. This direction was defined as 'vertical', meaning the lift was measured in a reference frame oriented with the water surface. The lift was measured while the model was towed at constant heel angles and during the dynamic roll tests. Figure 2 shows the lift force as a function of heel angle at three model speeds during the steady heel tests for two displacements. The lift force for the heavier displacement shows no speed dependence, although the lift force increases with speed for the lighter displacement. The heavier displacement steady lift force shows little dependence on heel angle until heeling beyond the deadrise angle of the hull. However, the steady lift force does depend on heel angle for the lighter displacement, especially at the faster model speeds. Therefore, there can be a coupling between heave and heel, at least when the vessel is running lightly loaded.

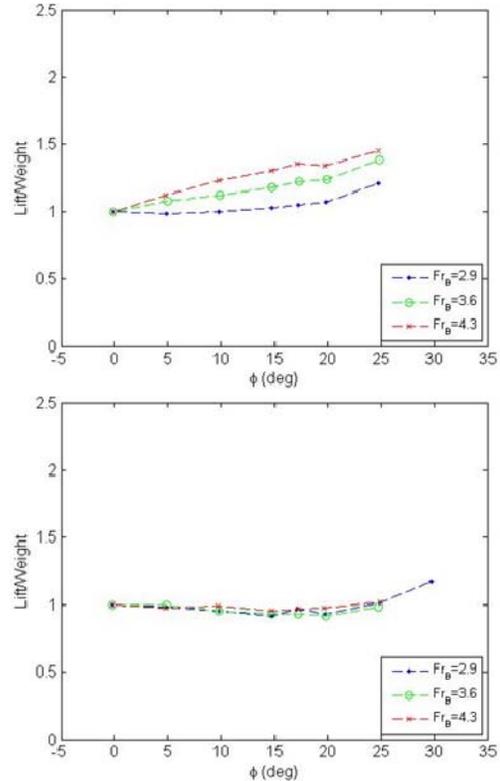

**Figure 13:** Lift force for steady roll angle model tests; lighter displacement (top) and heavier displacement (bottom).

Wetted Lengths

The keel wetted length is defined as the distance from the transom along the keel to where the keel is first wetted. The chine wetted lengths are defined as the distance from the transom along the chine to where the chine is first wetted. Figure 14 shows an underwater photograph of the model with the wetted lengths labeled.

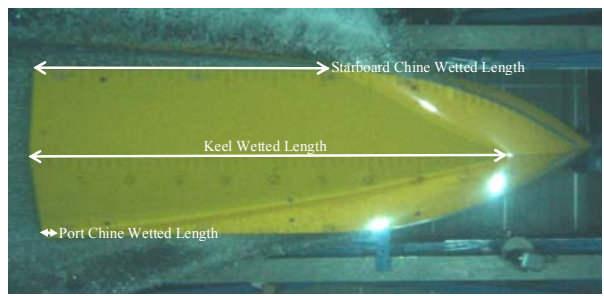

**Figure 14:** Underwater photograph of the model showing the different wetted lengths ($Fr_B = 4.3$, roll angle = 10º).

Savitsky (1964) provides wetted length predictions for steady planing hulls with no roll angle. However, there is no currently available method for predicting wetted lengths for hulls experiencing roll, even for steady planing. Wetted length is important for determining the hydrodynamic forces on a planing hull. Therefore, it is critical to understand the difference in wetted length due to roll angle or dynamic motion.

Figures 15 through 17 show the keel, starboard chine, and port chine wetted lengths for three different model speeds as a function of roll angle for the steady heel tests. The keel wetted length does not vary significantly with heel angle, while the port and chine wetted lengths show opposite trends. As would be expected, the starboard and port chine wetted lengths are close to equal at zero heel. As the model heels to starboard, the starboard chine wetted length increases and the port chine wetted length decreases. Eventually the starboard chine wetted length equals the keel wetted length and the port wetted length becomes zero. When the model is heeled to 20 degrees, the bottom of the model is flat from the keel to the chine since the deadrise is 20 degrees. For this condition, it is to be expected that the starboard and keel wetted lengths are equal. This is the case for the lighter displacement, but for the heavier displacement scenario shows the keel wetted length still greater than the chine wetted.

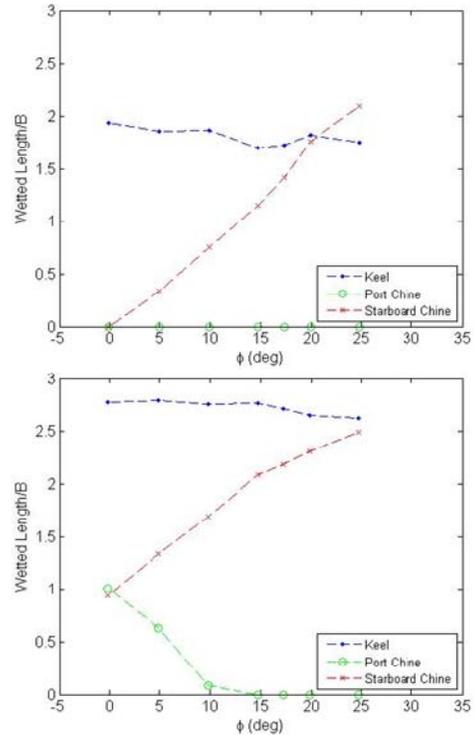

**Figure 16:** Steady Heel – Keel, starboard and port chine wetted lengths for model speed $Fr_B$ = 3.6; lighter displacement (top) and heavier displacement (bottom).

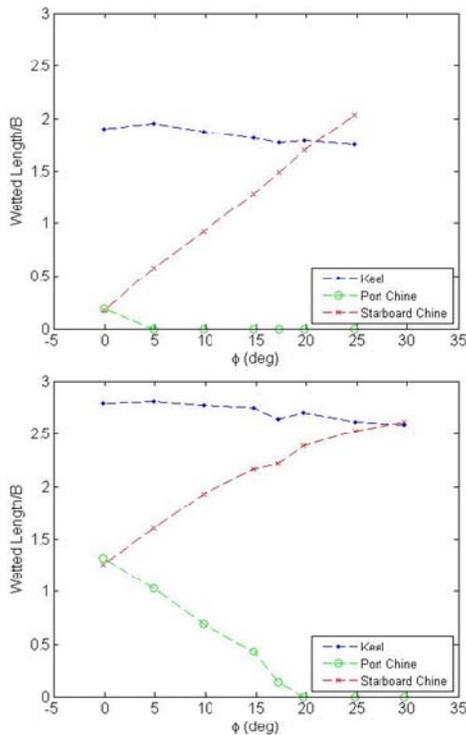

**Figure 15:** Steady Heel – Keel, starboard and port chine wetted lengths for model speed $Fr_B$ = 2.9; lighter displacement (top) and heavier displacement (bottom).

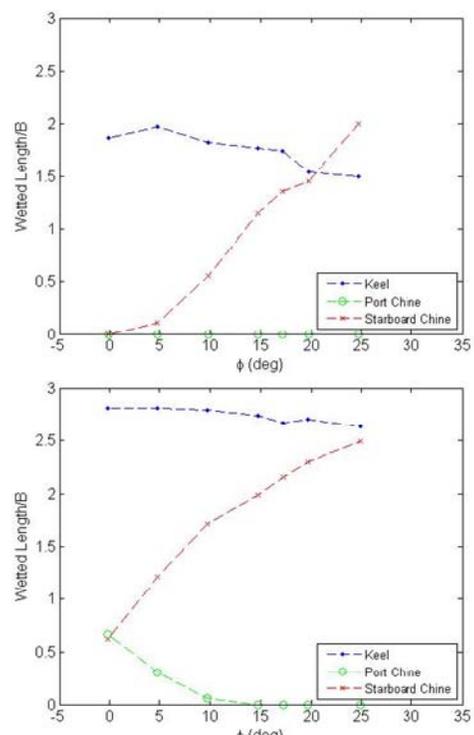

**Figure 17:** Steady Heel – Keel, starboard and port chine wetted lengths for model speed $Fr_B$ = 4.3; lighter displacement (top) and heavier displacement (bottom).

The keel wetted length shows a slight decrease as steady roll angle increases. This is likely because the model was fixed in heave and rotated about an axis through the model's center of gravity. As the roll angle becomes large, the keel moves up slightly resulting in a decrease in keel wetted length. The port wetted length is zero if the spray reaches the transom before reaching the chine. Figure 18 shows a photograph demonstrating zero chine wetted lengths.

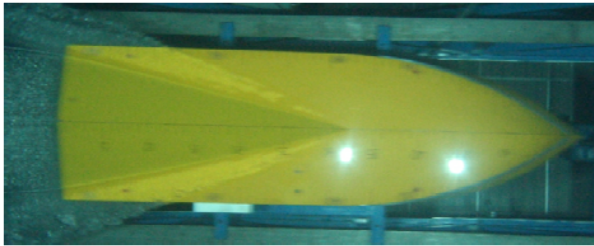

**Figure 18:** Underwater photograph of the model showing zero chine wetted lengths ($Fr_B$=4.3, zero roll angle).

The wetted lengths can be different for the same roll angle when the model is experiencing dynamic roll. As the model rolls to starboard, it would be expected that the starboard chine wetted length would increase compared to the same angle at zero roll velocity and the port chine wetted length would decrease comparatively. The keel wetted length should not be significantly affected by the roll motion. Figures 19 and 20 show the keel and starboard chine wetted lengths as a function of roll angle for the steady and dynamic roll angles for a single model speed. The dynamic wetted lengths include measurements from all roll amplitude and roll oscillation frequencies tested. As expected, the keel wetted length is relatively constant for both the steady and dynamic tests. The dynamic motion of the model does not significantly affect the wetted portion of the keel. The chine wetted lengths also show little variation due to dynamic roll motion.

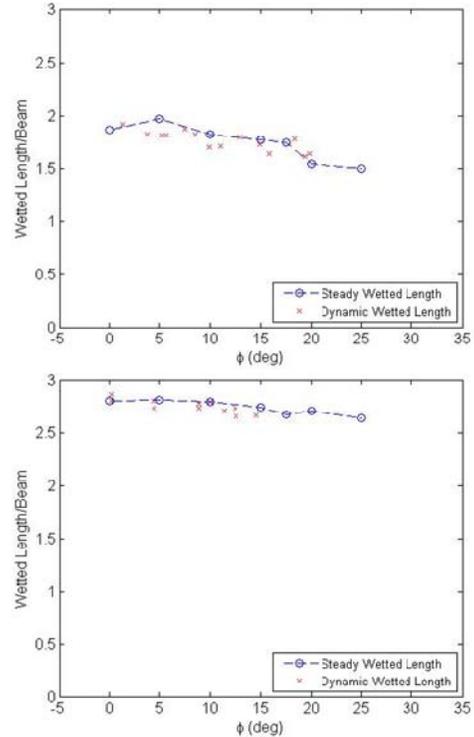

**Figure 19:** Steady and dynamic roll keel wetted Lengths for model speed $Fr_B = 4.3$; lighter displacement (top) and heavier displacement (bottom).

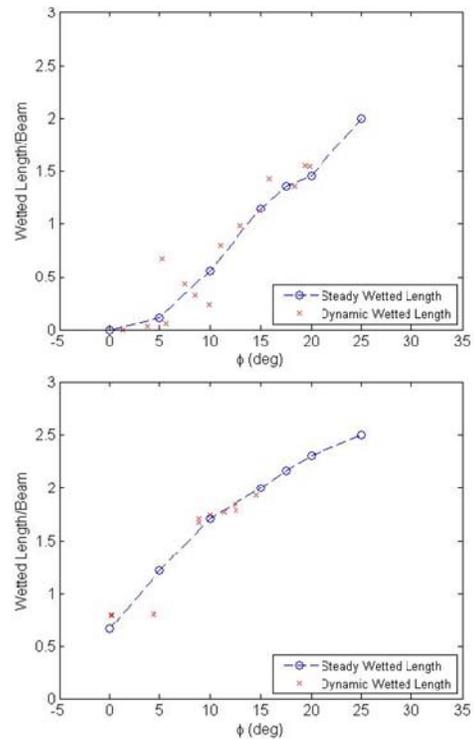

**Figure 20:** Steady and dynamic roll starboard wetted lengths for model speed $Fr_B = 4.3$; lighter displacement (top) and heavier displacement (bottom).

**NFA Comparison**

The NFA validation effort focused on the steady roll portion of the test, that is, the NFA simulations were performed only for the steady roll cases. The USNA geometry and NFA's domain can be seen in Figure 21. This constant deadrise hullform had an overall length of 1.52 m (5 ft) and a beam of 0.45 m (1.48 ft). Four cases of varying roll angles (0, 10, 20, and 30 degrees) at 6.1 m/s (20 ft/s) were investigated. The geometry was positioned at a trim of 4 degrees and a draft at the transom of 9.08 cm (3.574 in), and then rolled around an axis parallel to the keel at the VCG of 13.44 cm (5.29 in). The simulation was run with a width of 2 boat lengths, or 3.05 m (10 ft) and a depth of 1 boat length, or 1.52 m (5 ft). NFA's domain had to be smaller than the towing tank to cluster cells near the body without stretching the grid too heavily. This smaller domain may have resulted in some blockage and further simulations need to be performed to investigate the effect of domain size. The number of cells in x, y, and z was 1280, 896, and 448, resulting in 514 million cells in the total simulation. Spacing near the body was 0.0009L or 0.137 cm (0.054 in), necessitating a nondimensional time step of 0.00023. Simulations were run for 13,000 time steps, or 3 body lengths, which was sufficient to reach steady state, on Garnet, a Cray XT6 at ERDC and took approximately 24 hours to complete using 576 processors.

Figures 22 through 25 show the underwater photographs taken during the experiments compared to a similar view of the NFA simulations. The spray root line and thus wetted surface area agrees well to the experiments. The generation of the spray sheet is correctly predicted; the spray edge matches well to experiments. NFA also models the instabilities in the spray sheet breakup reasonably well. As the model rolls over the flow over the chine initially separates cleanly, but at 20 degrees and higher the flow wraps around the chine and wets the side of the model. The NFA simulations exhibit this behavior as well.

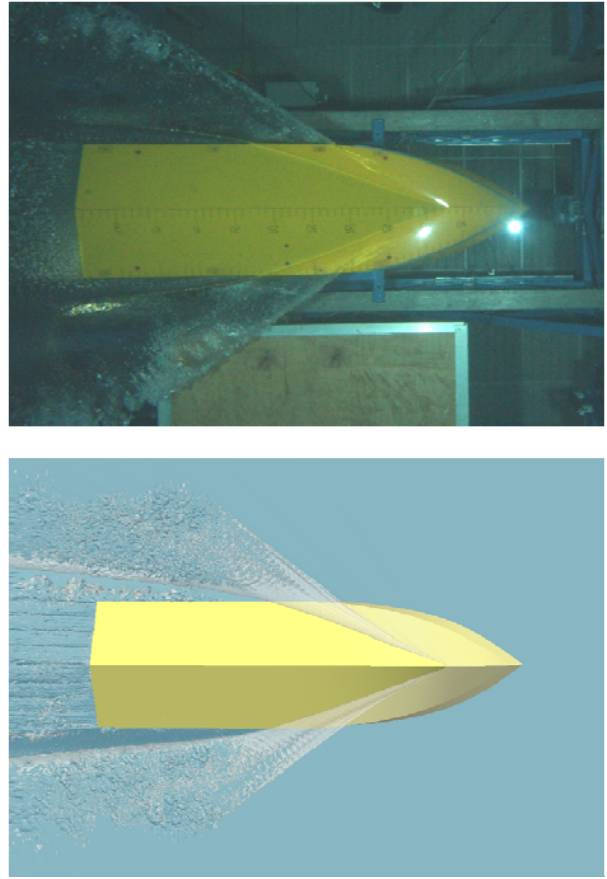

**Figure 22:** USNA underwater photograph compared to NFA Simulation; 0 degrees roll, U=20 ft/sec.

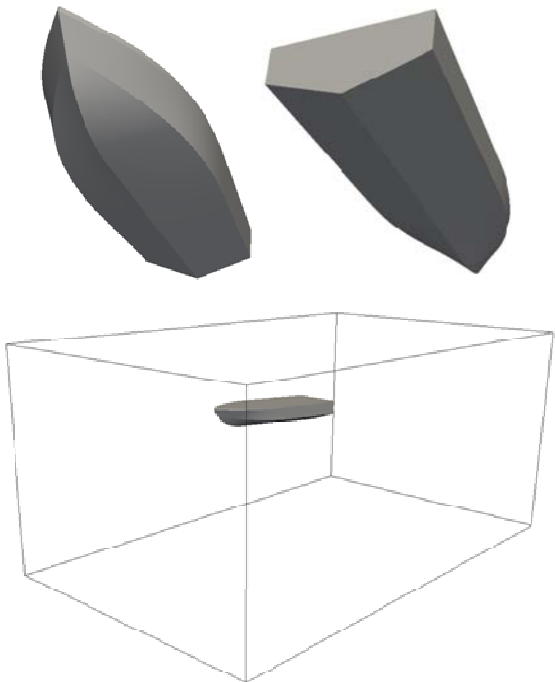

**Figure 21:** USNA Constant deadrise hullform (top) and NFA setup with geometry and domain extents (bottom).

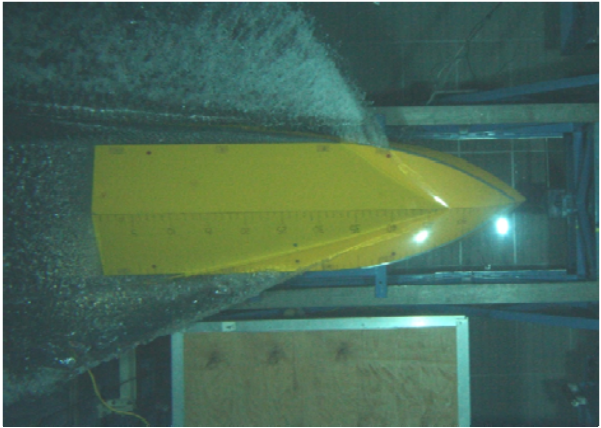
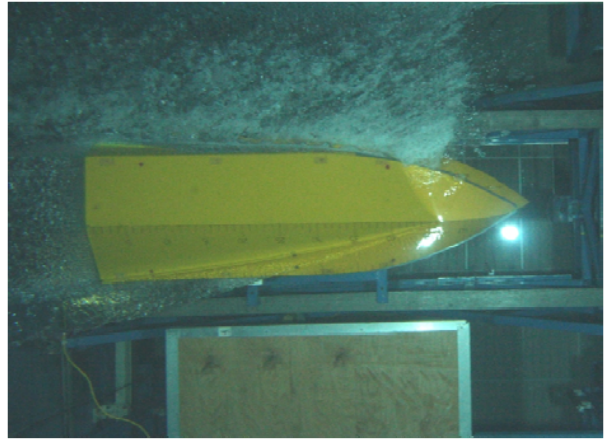
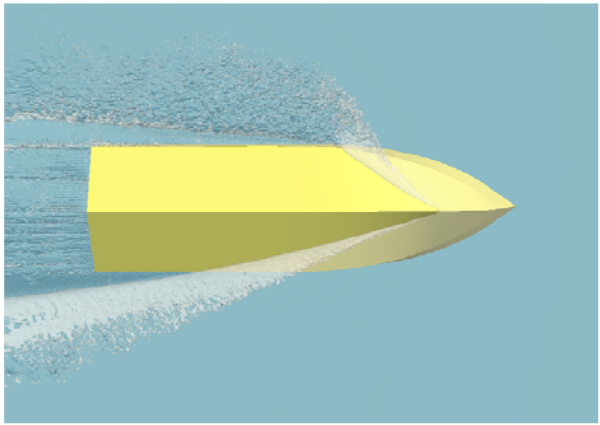
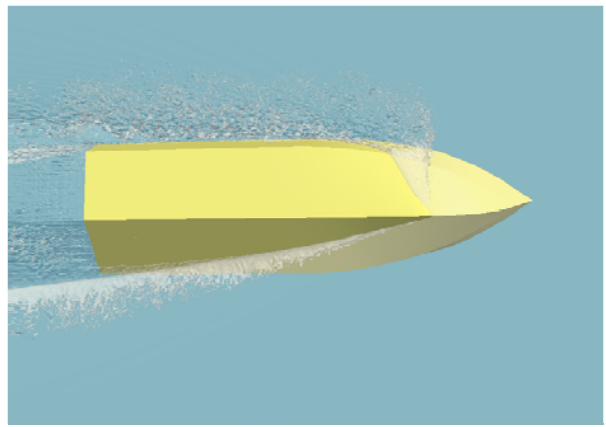

**Figure 23**: USNA underwater photograph compared to NFA simulation; 10 degrees roll, U=20 ft/sec.

**Figure 24:** USNA underwater photograph compared to NFA simulation; 20 degrees roll, U=20 ft/sec.

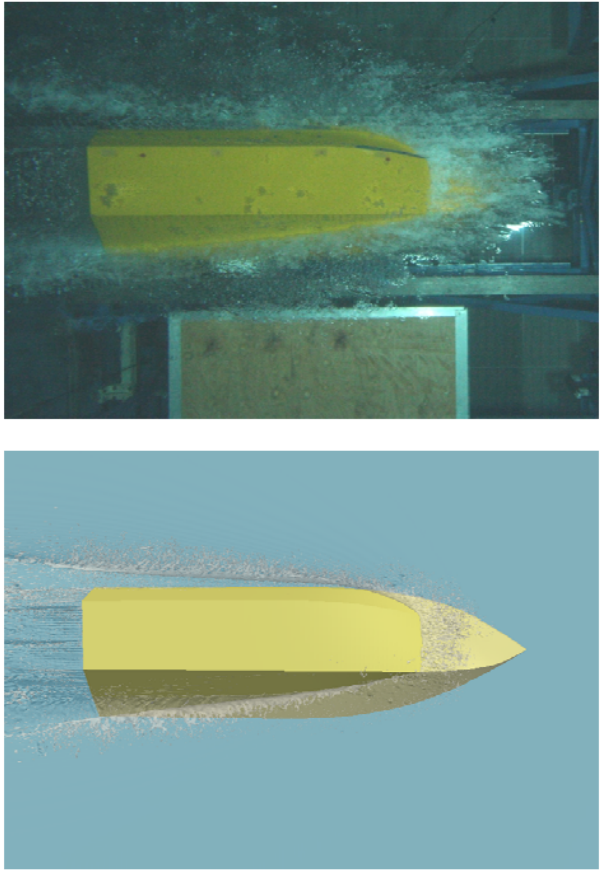

**Figure 25:** USNA underwater photograph compared to NFA simulation; 30 degrees roll, U=20 ft/sec.

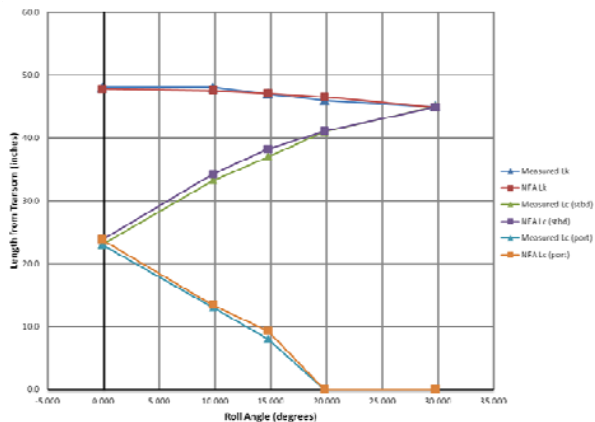

**Figure 26:** Comparison of NFA results to experiments for length along the keel, length along the starboard chine and length along the port chine for each roll angle simulated. Lengths are measured in inches from the transom.

The bottom of the model was painted with a ruler and the length along the keel and chines from the transom was estimated from the underwater pictures by the experimentalist. The same values were extracted from the NFA results and the comparison is plotted as Figure 26. The agreement is excellent throughout the range of roll angles. It follows that wetted surface area is being accurately predicted as well.

## SUMMARY

A recent effort to assess NFA's current ability to predict the hydrodynamic forces and moments of a deep-V planing craft and how well it models the complex multiphase flows associated with high Froude number flows, utilized data only form the classic planing boat work performed by Fridsma and Savitsky, but also recent experiments performed by Lesar at NSWCCD and Judge at the United States Naval Academy. This detailed validation effort was composed of three parts. The first part focused on assessing NFA's ability to predict pressures on the surface of a 10 degree deadrise wedge during impact with an undisturbed free surface. Detailed comparisons to pressure gauges are presented here for two different drop heights, 15.24 cm (6 in) and 25.4 cm (10 in). Results show NFA accurately predicted pressures during the slamming event. The second part examines NFA's ability to match sinkage, trim and resistance from Fridsma's experiments performed on constant deadrise planing hulls (Fridsma, 1969). Simulations were performed on two 20 degree deadrise hullforms of varying length to beam ratios (4 and 5) over a range of speed-length ratios (2, 3, 4, 5, and 6). Results show good agreement with experimentally measured values, as well as values calculated using Savitsky's parametric equations (Savitsky, 1964). The final part of the validation study focused on assessing how well NFA was able to accurately model the complex multiphase flow associated with high Froude number flows, specifically the formation of the spray sheet. NFA simulations of a planing hull fixed at various angles of roll (0, 10, 20, and 30 degrees) were compared to the USNA experiments. Comparisons to underwater photographs illustrated NFA's ability to model the formation of the spray sheet and the free surface turbulence associated with planing boat hydrodynamics. Overall these three studies demonstrate NFA's current capabilities to predict the hydrodynamics of a deep-V planing hull. NFA is capable of being a useful tool for evaluating novel planing hull-forms.

**ACKNOWLEDGEMENTS**

The authors would like to acknowledge Richard for providing the high quality wedge impact data used in this paper.

This work is supported by the Office of Naval Research (ONR), program manager Bob Brizzolara. The numerical simulations have been performed on the Cray XE6 platforms located at the U.S. Army Engineering Research and Development Center (ERDC) and the Air Force Research Laboratory (AFRL). NFA research and development has been sponsored over the years by Dr. Steve Russell and Dr. L. Patrick Purtell. SAIC IR&D also supported the development of NFA. This work is also supported in part by a grant of computer time from the Dept. of Defense High Performance Computing Modernization Program, http://www.hpcmo.hpc.mil/. Animations of NFA simulations are available at
http://www.youtube.com/waveanimations.